\documentclass[preprint,12pt]{elsarticle}

\usepackage{lineno,hyperref}
\modulolinenumbers[5]

\usepackage{subfigure}
\usepackage{float}

\usepackage[T1]{fontenc}

\usepackage{graphicx}
\usepackage{wrapfig}
\usepackage{amsmath}
\usepackage{amsfonts}
\usepackage{amssymb}

\begin{document}

\markboth{Jiying Jia, Dieter W. Heermann}
{Fractality and Topology of Self-Avoiding Walks}

\begin{frontmatter}

\title{Fractality and topology of self-avoiding walks}

\author{Jiying Jia}

\address{Institute for Theoretical Physics, Heidelberg University, Philosophenweg 19\\
Heidelberg, D-69120,
Germany}

\author{Dieter W. Heermann}
\ead{heermann@tphys.uni-heidelberg.de}
\address{Institute for Theoretical Physics, Heidelberg University, Philosophenweg 19\\
Heidelberg, D-69120,
Germany}

\begin{abstract}
We have analyzed geometric and topological features of self-avoiding walks. We introduce a new kind of walk: the loop-deleted self-avoiding walk (LDSAW) motivated by
the interaction of chromatin with the nuclear lamina. Its critical exponent is calculated and found to be different from that of the ordinary SAW. Taking the walks as point-clouds, the LDSAW is a subset of the SAW. We study the difference between the LDSAW and SAW by comparing their fractal dimensions and growth rates of the Betti number. In addition, the spatial distribution of the contacts inside a SAW, which is also a subset of SAW, is analyzed following the same routine. The results show that the contact-cloud has a multi-fractal property and different growth rates for the Betti number. Finally, for comparison, we have analyzed random subsets of the SAW, showing them to have the same fractal dimension as the SAW.
\end{abstract}


\end{frontmatter}

\section{Introduction}
\label{FracTopsec:intro}
One important feature in the organization of a  chromosome is the formation of the lamina-associated domains (LADs), where chromatin interacts with the nuclear lamina. These genome-lamina contacts are closely related to gene activities and have been explored intensively~\cite{kind2015a,vansteensel2017,kinney2018}. This type of interaction of chromatin with the surface can be viewed
as a polymer where those parts that are away from the lamina surface are considered loops. Now, consider the parts that are attached to the surface as a new polymer so that we have a surface polymer by erasing the loops of the original chromatin. This operation of loop erasure from the three dimensional polymer can be easily generalized to polymer chains in free space, where the loop is explicitly defined when two nonconsecutive monomers come into contact, that is, their distance is smaller than a cut-off value. 

The idea of loop erasure was first introduced by Lawler~\cite{lawler1980,lawler1988} where a different kind of
self-avoiding walk was defined by erasing the loops in a random walk, namely the loop-erased random walk (LERW). Due to its close 
relation to the uniform spanning tree and the Laplacian random walk, the LERW has received some 
attention since its introduction~\cite{marchal2000,lawler2006,schramm2011}. The LERW is self-avoiding but belongs to a different universality 
class from the normal SAW. It has a slightly larger critical exponent $\nu_\text{LERW} \approx 0.616$ in the 3D case which was 
intensively estimated in~\cite{guttmann1990}--\cite{agrawal2001}. Similarly, we wonder what the situation is if we erase the loops of a normal self-avoiding walk described above, specifically, whether the new walk has a different critical exponent.
To distinguish from LERW, we denote this new kind of walk as 
loop-deleted self-avoiding walk (LDSAW). 

A $N$-step self-avoiding walk can be viewed as $N+1$ non-overlapping monomers connected by $N$ bonds. If we ignore the connectivity, then it is a point cloud in space. The loop erasure procedure is to delete certain points from the cloud. The question is whether and how this deletion would change the geometrical and topological properties of the point cloud. We will explore this by calculating two kinds of indices. First is the fractal dimension which can measure the complexity of a shape. For the self-avoiding walk, the fractality is defined by $D_F=1/\nu$ ($\nu$ is the critical exponent) ~\cite{havlin1982,havlin1982a}. 
This relation should hold for the LDSAW, which we will show in the results section. 

Recently, owing to the powerful ability of persistent homology to reveal the underlying topological 
features of data at different resolutions~\cite{Bubenik2006,Fugacci2016}, the application of persistent homology is 
quite prevailing in many areas such as machine learning~\cite{Hofer2017,Giansiracusa2017}, disease 
identification~\cite{Nanda2014,im2016}, brain network~\cite{lee2012} and so forth. Specifically, the fractal property of 
data is linked to the persistent homology as they both deal with the data at different scales. Robins discussed the growth 
rates of the Betti number, i.e., the number of connected components or holes, as the resolution goes to zero, and explained the relation between 
these growth rates and the fractal dimension in her PhD thesis~\cite{Robins:2000}. Afterwards, Macpherson and 
Schweinhart~\cite{MacPherson2012} defined a persistent homology dimension based on the birth and death of holes to 
measure the complexity of the data. Another method to define the fractal dimension was introduced in~\cite{Mate2014}. 
Further, an estimator of fractal dimension in terms of the minimum spanning trees and higher dimensional persistent 
homology was proposed in~\cite{Adams2018}--\cite{Schweinhart2018a}. Despite all these different definitions, the 
fractal dimension established by persistent homology gives a description of the complexity of the system under 
investigation. In this paper, we utilize this tool to analyze the LDSAW.

As mentioned, each loop corresponds to a contact of two nonconsecutive monomers.
The contacts of polymers, especially of biomolecules, often play important roles in associated
systems. For example, certain biological functions are accomplished via contact and 
interaction of 
different parts of the molecule~\cite{kagey2010}. Experimentally, Hi-C contact maps are used to study the 3D 
organization of genome. The 
scaling behavior of the contact probability of a self-avoiding walk (SAW) as a function of its
contour length was studied in~\cite{des1980short}--\cite{jia2019effect}. The average number of contacts 
$\langle m \rangle$ of SAW is found to have an asymptotic behavior $\langle m \rangle \sim a_{\infty} N$, where $N$ is 
the length of the SAW~\cite{Douglas1995}. Baiesi~\cite{baiesi2001peculiar} studied the contacts between two halves of a 
SAW and found that they are strictly finite in number. The dependence of the number of contacts on the radius of 
gyration of a SAW also has a simple scaling law~\cite{victor1994a}. In this paper, we would rather inspect the
spatial 
distribution of these contacts. Likewise, these contacts are a subset of the point cloud of a SAW and we will also study its fractality and topology. 

In this paper we first give the definition of the loop-deleted self-avoiding walk (LDSAW) and two ways to estimate its critical exponent in section~\ref{sec:ldsawdef}. 
Then, we introduce several definitions of the fractal dimension (section~\ref{FDdefs}), one of which is calculated using the Barycentric fixed-mass method~\cite{Kamer2013}.
In section~\ref{PHintro}, the basic idea of the persistent homology and the growth rate of the Betti number are illustrated. 
In section~\ref{FracTopsec:results} we present our results. The fractal dimension and the growth rates of Betti number of the self-avoiding walk are estimated in section~\ref{subsec:FDSAW}. The critical exponent of the loop-deleted self-avoiding walk, as well as the fractal dimension and growth rates, are calculated in section~\ref{sec:LDSAW}. The spatial distribution of the contacts inside a self-avoiding walk is analyzed both geometrically and topologically in section~\ref{sec:contact}. 
We find that the contact-cloud has a multi-fractal property. 
Both the LDSAW and the contact-cloud are subsets of point-cloud of the self-avoiding walk, yet they have different fractal dimension and topological indexes. 
For comparison, in section~\ref{sec:rdsaw} we study the random set of the SAW point-cloud. Finally, in section~\ref{conclu} we present conclusions.

\section{Concepts and Methods}

\subsection{The Loop-Deleted Self-Avoiding Walk}
\label{sec:ldsawdef}
\begin{figure}[bt]\centering
\subfigure[]{\includegraphics[width=0.35\textwidth]{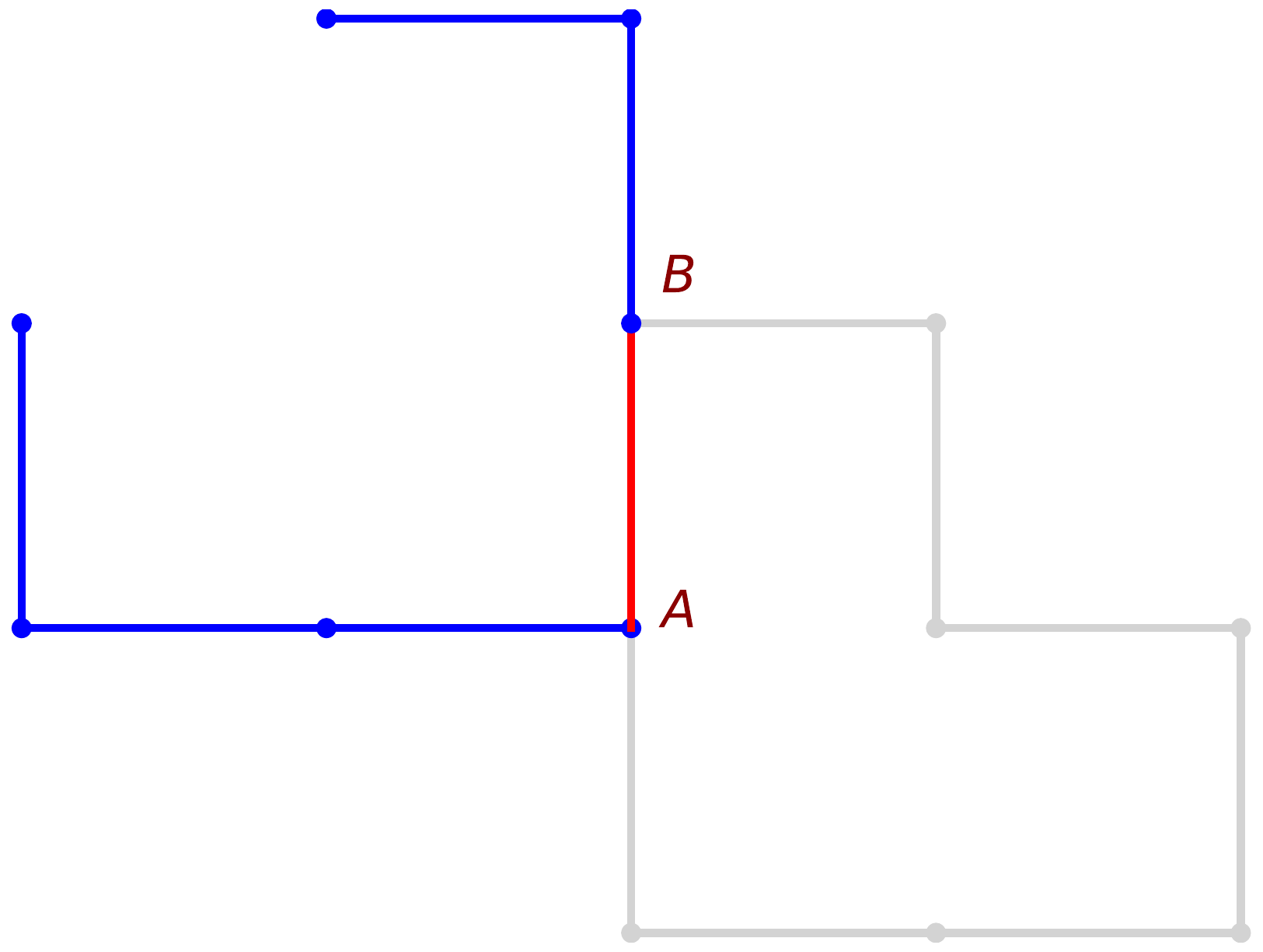}}
\subfigure[]{\includegraphics[width=0.45\textwidth]{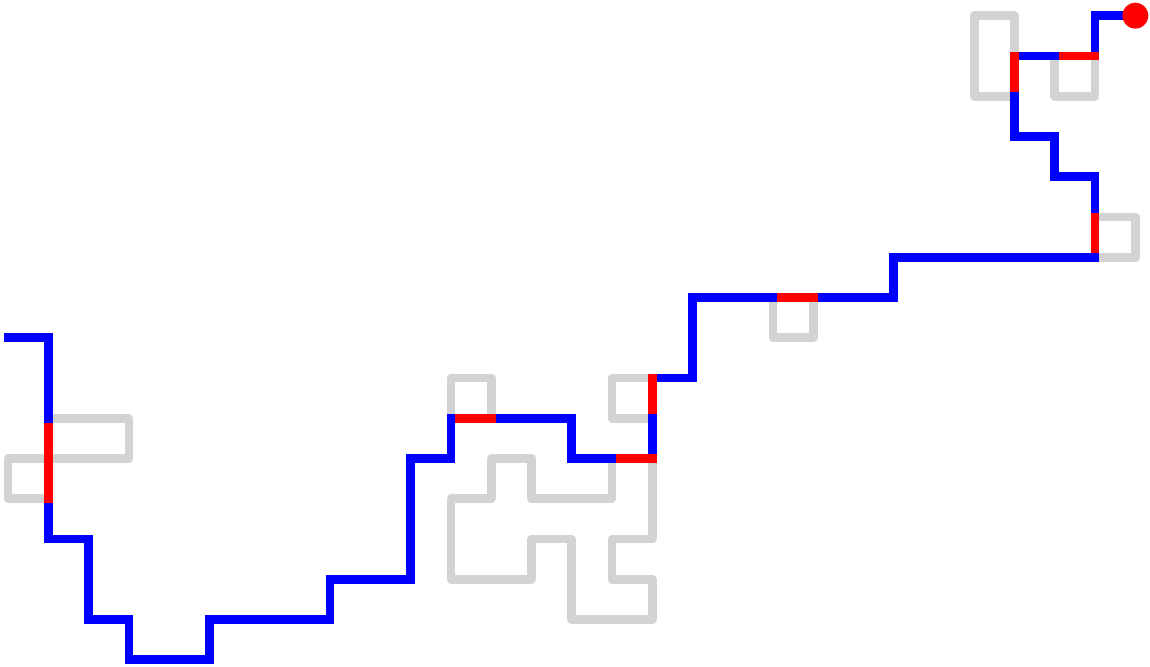}}
\caption{The loop deletion of a 2D self-avoiding walk on a square lattice. (a) The nonconsecutive monomer $A$ and $B$ of 
SAW come into contact and a loop arises. By deleting all the monomers between $A$ and $B$ (the gray part) and 
connecting $A, B$, we get the LDSAW. Shown in (b) is an example of a longer LDSAW. The loops are detected and deleted from the end on the right side (the red point).}
\label{fig:ldsaw}
\end{figure}

The definition of the loop-deleted self-avoiding walk is quite similar to the loop-erased random walk. Let $d(.,.)$ denote a distance measure.
Suppose we have a $N$-step self-avoiding walk on a lattice denoted by $S_N=\{r(i):i=0,1,...,N\}$, where $r(i)$ is the position of the $i$th monomer, $d(r(i),r(i+1))=1$, and $r(i)\neq r(j)$ for $i\neq j$. The new walk LD$(S_N)$ is obtained by deleting the loops of $S_N$ in chronological order. Starting from one end of the walk, if the $k$th monomer of LD$(S_N)$ corresponds to the $i_k$th monomer $r(i_k)$ of $S_N$, then
\begin{equation}
    i_0=0, i_{k+1}=\max\{i:d(r(i),r(i_k))=1\}.
\end{equation}
The schematic procedure of this loop deletion is shown in FIG.~\ref{fig:ldsaw}, where the gray parts denote the loops, and the red segments are newly created bonds in the LDSAW.

Clearly, the LDSAW is also self-avoiding. The difference is that the minimal distance between 
nonconsecutive monomers is $\sqrt{2}$ for the walk on a square or cubic lattice. 

It has already been shown that the loop-erased random walk (LERW) has a different critical exponent from the normal SAW: 
$\nu_{\text{LERW}}\approx0.616$~\cite{guttmann1990,bradley1995,wilson2010,agrawal2001}. Similarly, we expect another 
critical exponent $\lambda$ for the LDSAW.
The critical exponent $\lambda$ can be calculated in two ways. First, the average length of the LDSAW $\langle L_N\rangle$ from deleting loops of $N$-step self-avoiding walks has a power law dependence on $N$:
\begin{equation}
\langle L_N \rangle \sim N^{1/2\mu},\ N\rightarrow\infty \; .
\label{eq:LN-N}
\end{equation}
Note that the loop deletion does not change the end-to-end distance of the walk, therefore,
\begin{equation}
\langle R_e^2 \rangle_{\text{LDSAW}} = \langle R_e^2 \rangle_{\text{SAW}} \sim N^{2 \nu}
\sim \langle L_N \rangle^{4 \nu \mu} \; ,
\end{equation}
where $\nu\approx 0.588$ is the critical exponent of the self-avoiding walk. This gives an approximation of the critical exponent of LDSAW: $\lambda=2\nu\mu$.  To compute the exponent,
the pivot algorithm~\cite{kennedy2002faster} is used to generate long self-avoiding walks to get a better estimation of the exponent.

The second way is to obtain sufficient samples of the $L$-step LDSAW and calculate the mean square end-to-end distance $\langle R_e^2(L)\rangle_{\text{LDSAW}}$, thus $\lambda$ can be extracted by $\langle R_e^2(L)\rangle_{\text{LDSAW}}\sim L^{2\lambda}$. For LERW, it is feasible to erase the loops of a growing random walk until the length of LERW reaches $L$~\cite{bradley1995}. However, this method is impractical to generate long LDSAW due to the attrition problem. Therefore, we still need to generate the $N$-step self-avoiding walks and then delete the loops following the routine described above. Adopting the idea from~\cite{guttmann1990}, for each LDSAW with $L_N\geqslant L$, the square distance from the 0th monomer to the $L$th monomer is recorded and contributes to $\langle R_e^2(L)\rangle_{\text{LDSAW}}$. $L$ should be much smaller than $N$ such that the samples of $L$-step LDSAW 
from deleting loops of SAW longer than $N$ account for a negligible part of the ensemble of $L$-step LDSAW. Thus, the mean square end-to-end distance of $L$-step LDSAW can be approximated with a very small sampling bias. 

\subsection{Definitions of Fractal Dimensions}
\label{FDdefs}
A self-avoiding walk can also be viewed as a point-cloud $\mathcal{S}$ if the connectivities between monomers are ignored. In this manner, the point-cloud $\mathcal{L}$ of LDSAW is actually the result of deleting some points in $\mathcal{S}$ following a certain routine. The question is whether and how the deletion changes the structure of the point-cloud $\mathcal{S}$. We shall study this from the geometrical and topological aspects. One important geometrical measure of a shape is its fractal dimension which describes how the shape is changing at different scales. The fractal dimension of the self-avoiding walk is the reciprocal of the critical exponent $D_F=1/\nu$~\cite{havlin1982,havlin1982a}.

There are several definitions of the fractal dimension, of which the box counting dimension $D_{\text{box}}$ is the simplest one. If $S$ is the 
fractal under study, 
\begin{equation}
D_{\text{box}} (S) = \lim_{l \rightarrow 0} \frac{\log N (l)}{\log (1 / l)}, 
\label{boxD}
\end{equation}
where $N(l)$ is the number of boxes of side length $l$ needed to cover $S$. An equivalent definition is given by 
\begin{equation}
 D_{\text{MB}} (S) = n - \lim_{l \rightarrow 0} \frac{\log \text{vol}
(S_l)}{\log l},
\end{equation}
where $\text{vol}(S_l)$ is the influence volume of $S$ by dilating $S$ by a sphere of radius 
$l$~\cite{dubuc1989,Reichert2017}. $n$ is the dimension of the Euclidean space. 

Another important definition of fractal 
dimension is the mass dimension, which is defined as the exponent of power law relation between the mass $M_{\mathbf{x}}(l)$ within 
a ball of radius $l$ centered at a point $\mathbf{x}$:
\begin{equation}
{D_{\text{mass}}}(\mathbf{x})=\lim_{l\rightarrow 0}\frac{\log M_{\mathbf{x}}(l)}{\log l}.
\end{equation}
The mass dimension is quite intuitive if we consider that for a $n(=1,2,3)$ dimensional object, $M_{\mathbf{x}}(l)$ is proportional 
to $l^n$~\cite{voss1986,peitgen1988}. Taking the limit $l\rightarrow 0$ and the averaging of $M_{\mathbf{x}}(l)$ over 
every point in different sequence gives the information dimension and the correlation dimension~\cite{theiler1990}. The box-counting dimension together with the 
information dimension and correlation dimension are special cases of the generalized dimension which is:
\begin{equation}
D_q = \lim_{l \rightarrow 0} \frac{\frac{1}{1 - q} \log \left( \sum_i p_i^q \right)}{\log (1 / l)} \; ,
\label{dq-size}
\end{equation}
where $p_i$ is the percentage of points within box $i$ and $q$ is the moment. For a monofractal, the value of $D_q$ will not 
change with $q$, while if an object is a combination of different fractals, then $D_q$ is decreasing with $q$, 
which characterizes the structure of this multi-fractal. Specifically, $D_\infty$ and $D_{-\infty}$ correspond to the most 
dense and least dense areas.

Instead of counting the points within a box of size $l$, Termonia and Alexandrowicz~\cite{termonia1983} defined the 
fractal dimension by finding the scaling behavior of the average radius $\langle 
R_m\rangle$ of $m$ nearest-neighbor points with $m$:
\begin{equation}
 \langle R_m\rangle \propto m^{1/D} \; .
 \label{FMD}
\end{equation}
Afterwards, it was proposed in~\cite{radii1985} that the Eq. (\ref{FMD}) can be extended to the moment of order 
$-\tau$:
\begin{equation}
 \langle R_m^{-\tau} \rangle \propto m^{-\tau / D (\tau)} \; ,
 \label{dq-mass}
\end{equation}
where $\tau = (q - 1) D_q, D (\tau) = D_q$.
The algorithms based on Eq.~(\ref{boxD}) and~(\ref{dq-size}) are fixed-size algorithms, while algorithms based on 
Eq.~(\ref{FMD}) and~(\ref{dq-mass}) are fixed-mass algorithms~\cite{theiler1990}. The latter outperforms the 
fix-sized algorithms in some cases, especially when $q<1$~\cite{grassberger1988,badii1988}.

By reducing the SAW or LDSAW to a point-cloud, the fractal dimension can be estimated by different algorithms based on these definitions. 
The Barycentric fixed-mass method~\cite{Kamer2013} is used here owing to its robustness, where the barycentric pivot point selection and non-overlapping coverage criteria are incorporated to reduce the finite size and edge effects.

\subsection{Persistent Homology}
\label{PHintro}
To explore the topological difference of the point-cloud after deleting certain points, we carried out a persistent homology analysis. The scheme of persistent homology is to discover how the topological features of a shape, such as the connected 
components, holes and voids, would change at different scales. The point-cloud that describes the 
shape needs to be represented by a simplicial complex. Then a filtration is started by growing the balls centered at 
each point. The radius of the balls, or the resolution is denoted by $\epsilon$. 
During this growing process, the number of the connected components is always non-increasing, while the holes and voids 
could appear and disappear when increasing $\epsilon$. The numbers of connected components, holes and voids of a shape are denoted as the 0th, 1st, 2nd 
persistent Betti number $\beta_i \; (i=0,1,2)$. The values of $\epsilon$ when the appearance and disappearance happen are denoted as 
$\epsilon_{birth}$ and $\epsilon_{death}$. The intervals $(\epsilon_{birth}, \epsilon_{death})$ represent the 
underlying topological property of the given shape, and can be visualized as the persistent barcodes and the persistent 
diagram. The long intervals, which means that the holes or voids persist through a large range of resolution, indicate 
important topological features, and very small intervals are often considered as noise. The persistent barcodes and diagrams have extensive applications in biology. 
For example, the 
barcodes are used to detect the existence of alpha helices and beta sheets in proteins~\cite{xia2014}. In addition, various definitions 
of distance between two persistent diagrams provide different measures for the similarity of two shapes, e.g. two 
$\gamma$H2AX clusters~\cite{hofmann2018}. Detailed mathematical description for the persistent homology can be found 
in~\cite{Fugacci2016} and \cite{otter2017}.

The connection of persistent homology with fractality was proposed by Robins~\cite{Robins:2000} where the growth rates 
of the Betti numbers are studied. For a shape that is fractal, suppose as $\epsilon\rightarrow 
0$, $\beta_i(\epsilon)\rightarrow\infty$, an exponent $\gamma_i$ could be defined if the asymptotic behavior is a power 
law:
\begin{equation}
\gamma_i = \lim_{\epsilon \rightarrow 0} \frac{\log \beta_i (\epsilon)}{\log
(1 / \epsilon)}.
\end{equation}
If the limit does not exist, alternatively limsup or liminf is used. By applying these to some well-defined fractals, Robins 
showed that the growth rates can distinguish sets with the same Hausdroff dimension but different homology. However, 
the relation between them and the fractal dimension 
remains an open question. 

Since a self-avoiding walk with 
equal bond length is already a minimal spanning tree, $\beta_0$ is $N$ or 1 for different $\epsilon$. The loops in the self-avoiding walk are 
actually one kind of holes in the first dimension of persistent homology.

\section{Results}
\label{FracTopsec:results}

\subsection{The Fractal Dimension of the 3D Self-Avoiding Walk}
\label{subsec:FDSAW}
\begin{figure}[H]\centering
\includegraphics[width=0.7\textwidth]{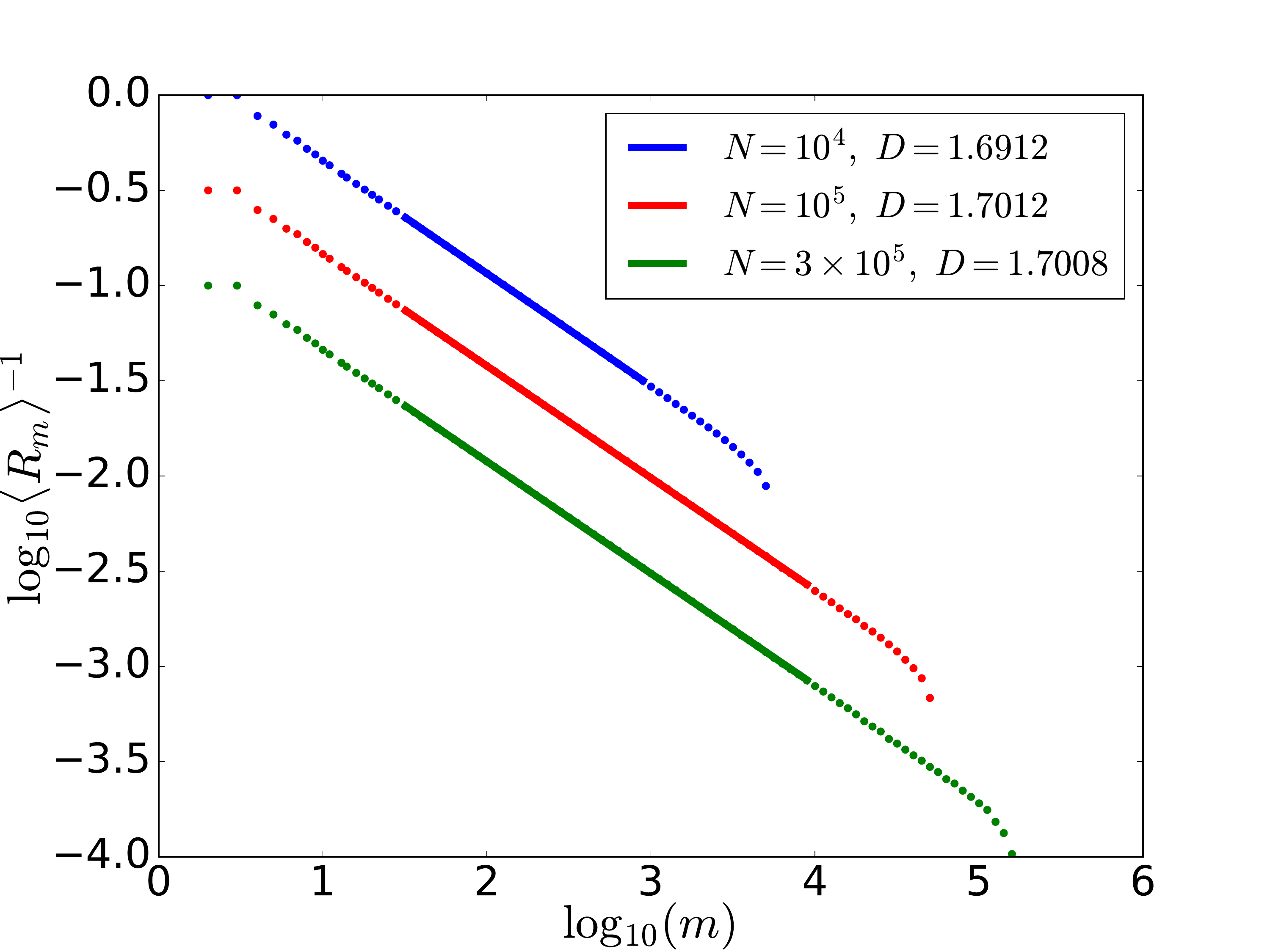}
\caption{Log-log plot of $\langle R_m\rangle$ versus $m$ for self-avoiding walks of $N=10^4, 10^5, 3\times 10^5$. The fractal dimension $D$ is related to the slope $a$ of the linear 
parts of the points by $D=-1/a$. The data of $N=10^5$ and $N=3\times 10^5$ are shifted downwards by 0.5 and 1 
intentionally to avoid overlapping.}
\label{fig:BFMSAW}
\end{figure}

The critical exponent for the 3D SAW is $\nu\approx0.588$, and the fractal dimension is
$D=1/\nu\approx 1.7$~\cite{havlin1982}. Usually $\nu$ is estimated by
the asymptotic behavior of mean-square end-to-end distance or radius of gyration: $\langle R_e^2\rangle\sim N^{2\nu}, 
\langle R_g^2\rangle\sim N^{2\nu}$, here we calculate the fractal dimension $D$ of the self-avoiding walk for $N=10^4, 10^5, 3\times 10^5$ using the Barycentric Fixed-Mass method~\cite{Kamer2013}.
Fig.~\ref{fig:BFMSAW}  shows the log-log plot of Eq.~\ref{FMD} for these walks, corresponding to $\tau=-1$ in Eq.~\ref{dq-mass}. 
The results are averaged over 15 000, 200, 100 conformations respectively. The data of $N=10^5$ (red dots) and $N=3\times 10^5$ (green dots) are shifted downwards intentionally to avoid overlapping. By fitting the linear regions, we get the estimated values of $D$: 1.6912, 1.7012, 1.7008, which are quite close to the expected value 1.7.

\begin{figure}[H]\centering
\includegraphics[width=0.7\textwidth]{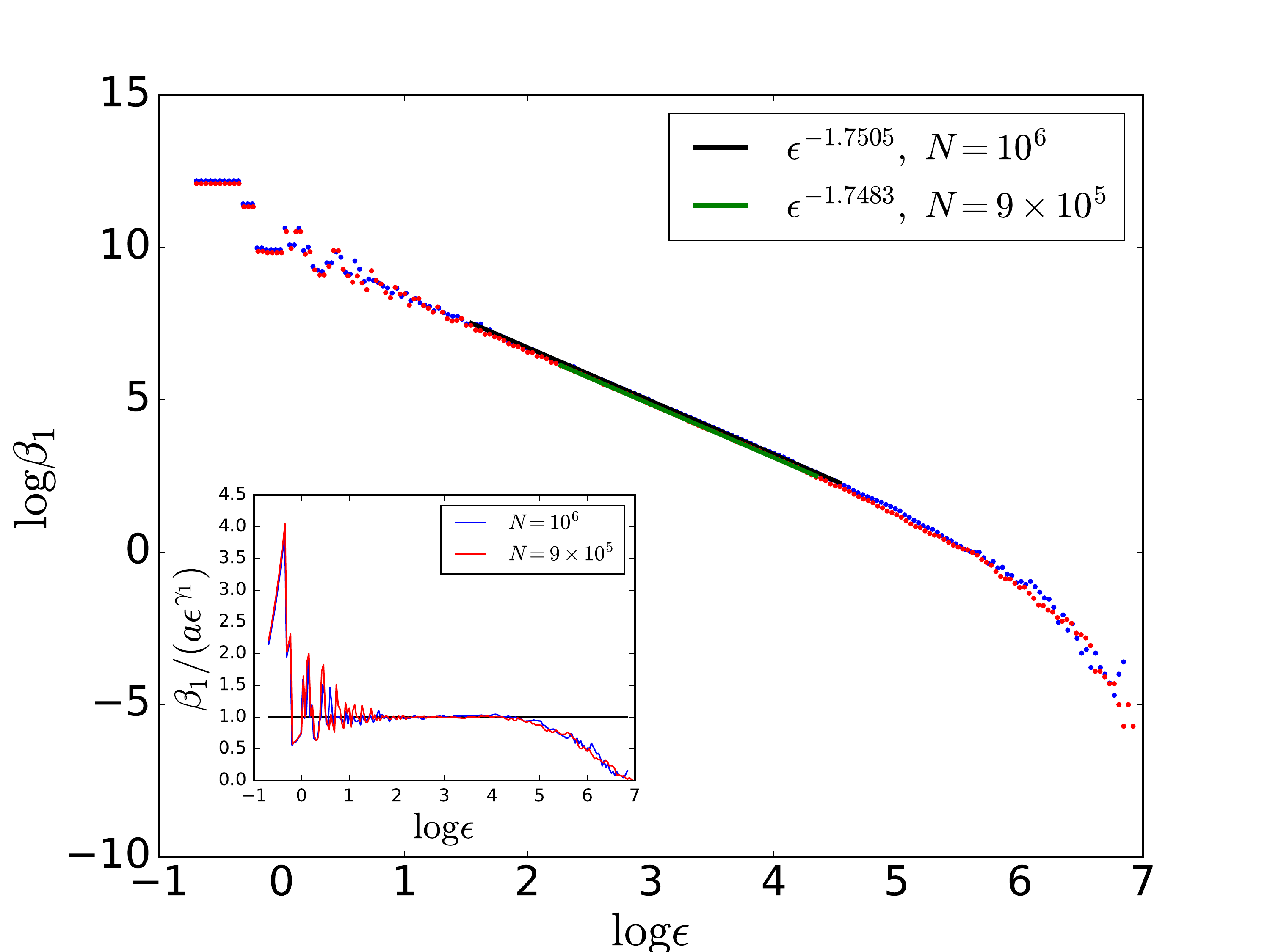}
\caption{Log-log plot of the dependence of the first Betti number $\beta_1$ on the resolution $\epsilon$ for
self-avoiding walks with $N=5\times 10^5$ and $N=10^6$.  The results are averages over $500$ and 
$300$ independent conformations. By fitting the linear regions of both set of points, we get the growth rate 
$\gamma_1\approx 1.75$. The inset shows the ratio between the original data and the power law relation.}
\label{fig:1stBettiSAW}
\end{figure}

The persistent homology of the self avoiding walk is calculated using the package Dionysus~\cite{dion:2020}, where the alpha 
complexes are constructed for the filtration due to its high efficiency when dealing with large systems. We analyze 
self-avoiding walks of length $N=9\times10^5$ and $N=10^6$. The first Betti number $\beta_1(\epsilon)$, which corresponds to the number of holes, is counted when increasing the resolution $\epsilon$.
500 and 300 independent conformations are 
generated to average $\beta_1(\epsilon)$ respectively. Fig.~\ref{fig:1stBettiSAW} shows the log-log plot of the first Betti number versus $\epsilon$. The growth rate $\gamma_1$ is estimated by fitting the linear region. The results are $\gamma_1\approx 1.7505$ for $N=10^6$ and 1.7483 for $N=9\times 10^5$.

\subsection{The Loop-Deleted Self-Avoiding Walk}
\label{sec:LDSAW}
\begin{figure}[H]\centering
\subfigure[]{\includegraphics[width=0.49\textwidth]{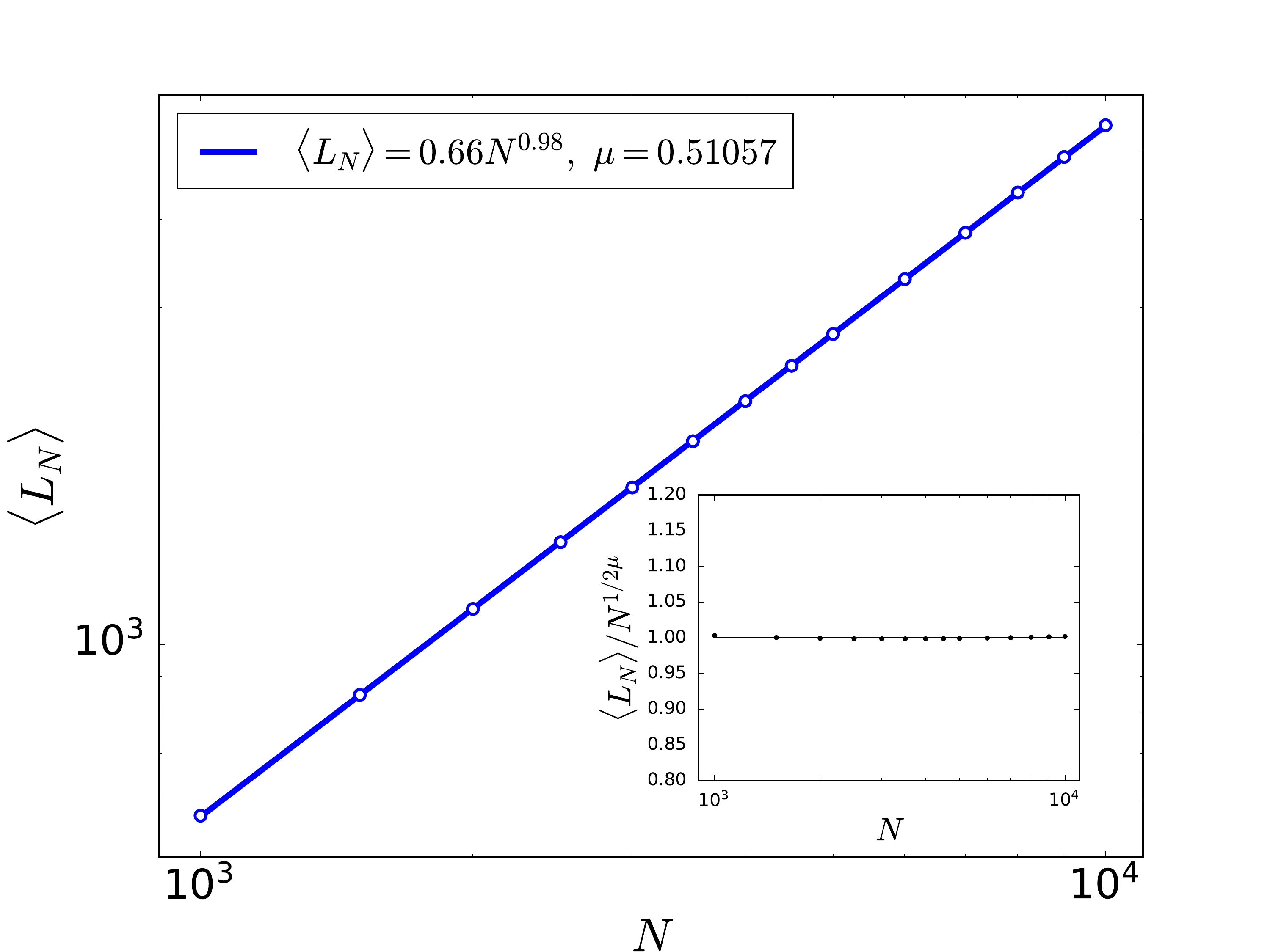}\label{mN-N-lesaw}}
\subfigure[]{\includegraphics[width=0.49\textwidth]{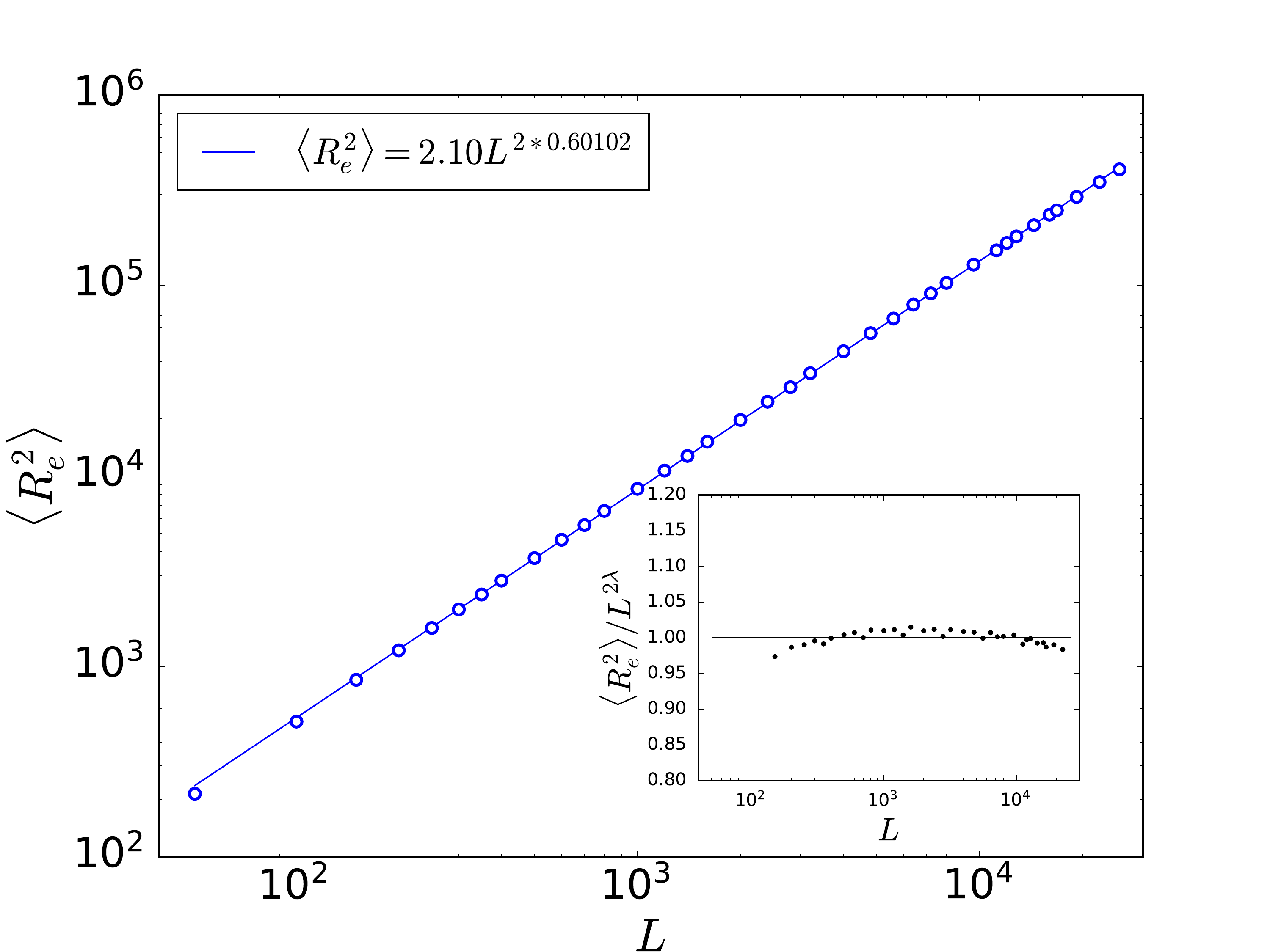}\label{re2m-lesaw}}
\caption{(a) Log-log plot of average length of LDSAW $\langle L_N \rangle$ versus $N$. Shown in the inset is the ratio 
between $\langle L_N \rangle$ and the power law by fit. (b) The dependence of the mean squared end-to-end distance $\langle R_e^2 \rangle$ of LDSAW on the length $L$. The 
inset is the ratio between $\langle R_e^2 \rangle$ and the power law.}
\end{figure}

The definition of the loop-deleted self-avoiding walk is given in section~\ref{sec:ldsawdef}. One important question for this walk is whether it has a different critical exponent. Above we mentioned two ways to calculate $\lambda$. 
First, Fig.~\ref{mN-N-lesaw} shows the relation between $\langle L_N \rangle$ and $N$ (see Eq.~(\ref{eq:LN-N})). By a linear fit in the log-log 
plot of $\langle L_N \rangle$ versus $N$, we get $\mu\approx 0.511$. Thus the estimated critical exponent for LDSAW is 
$\lambda=2\mu\nu\approx 0.600\pm 0.0004$.

The second way is to calculate the mean squared end-to-end distance of $L$-step LDSAW. Note that it is vitally important to get an unbiased sampling of the walk, which is detailed in section~\ref{sec:ldsawdef}. Fig.~\ref{re2m-lesaw} shows the dependence of the mean squared end-to-end distance on the length $L$. By linear regression, we get the exponent $\lambda\approx 0.601\pm 0.0006$.

These two methods 
give similar values of $\lambda$, therefore, we conclude that the loop-deleted self-avoiding walk has a critical exponent around $0.60$. This new 
exponent is larger than the exponent $\nu$ of SAW, which is reasonable since the loop deletion basically stretches the 
self-avoiding walk. 

Another point worth mentioning is that 
although the LDSAW is more stretched, it has a smaller critical exponent than the LERW (0.616). The reason is that the 
stretching of LDSAW is more local compared with the LERW.

\begin{figure}[H]\centering
\includegraphics[width=0.7\textwidth]{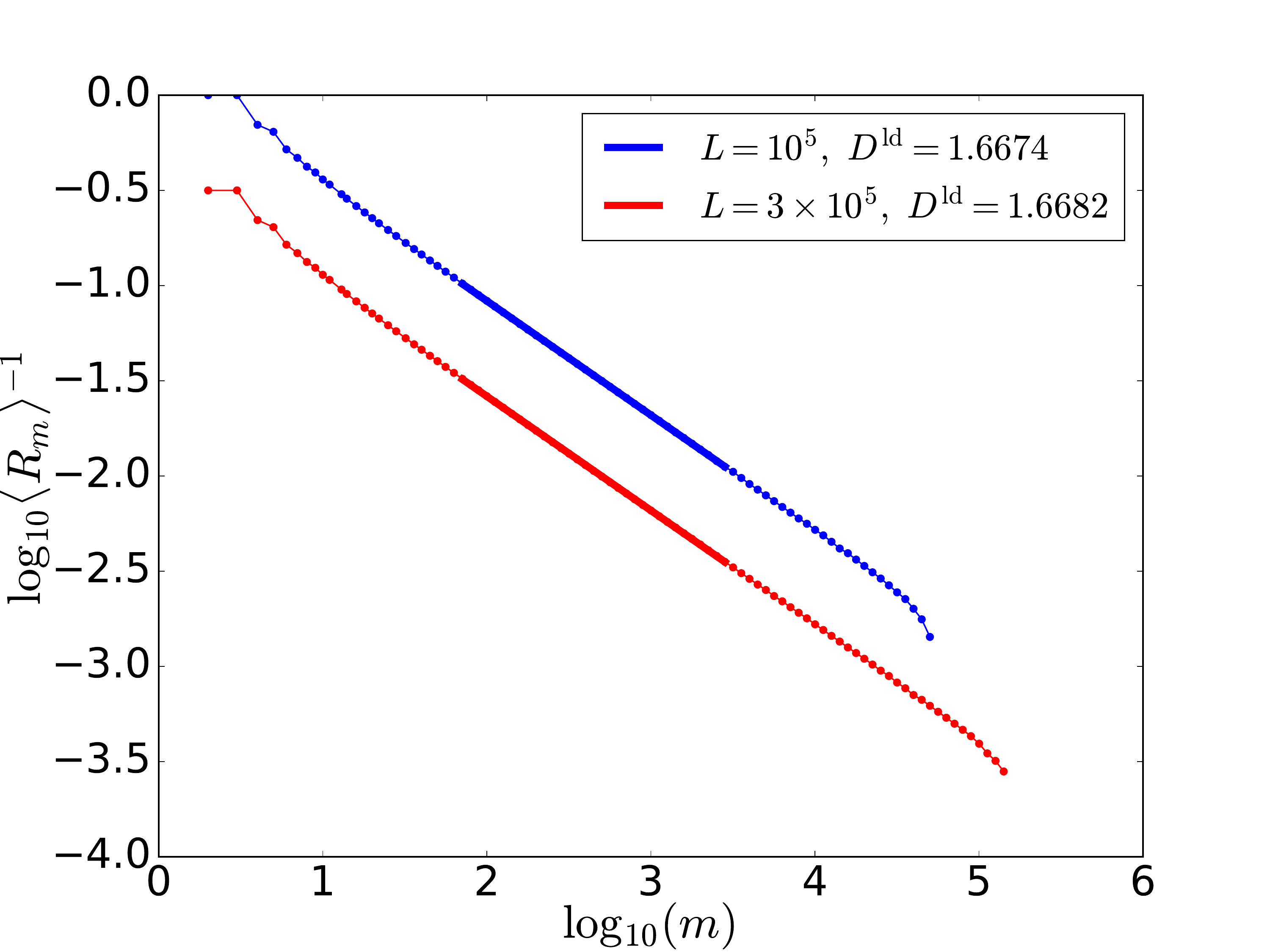}
\caption{Log-log plot of $\langle R_m\rangle$ versus $m$. The 
lengths of the LDSAW are $L=10^5$ and $L=3\times 10^5$. By determining the slope $a$ of the linear parts, the fractal 
dimension of LDSAW is $D^{\text{ld}}=-1/a\approx 1.6674$ and $1.6682$. The data of $L=3\times 10^5$ are shifted downwards by 0.5 intentionally to avoid overlapping.}
\label{fig:BaryDLDSAW}
\end{figure}

The fractal dimension of the LDSAW can be inferred by $D^{\text{ld}}=1/\lambda\approx 1.667$. To 
verify this, the Barycentric fixed-mass method is applied to the LDSAW with length $L=10^5, 3\times 
10^5$. Fig.~\ref{fig:BaryDLDSAW} shows the log-log plot of $\langle R_m\rangle$ versus the mass $m$ 
(Eq.~(\ref{FMD})). The results are averaged over 200 conformations. The fractal dimension can be extracted by fitting the linear regions of the points in the log-log plot. The estimated values of $D^{\text{ld}}$ for $L=10^5, 3\times 10^5$ are 1.6674 and 1.6682, which are quite close to the predicted value.

\begin{figure}[H]\centering
\includegraphics[width=0.7\textwidth]{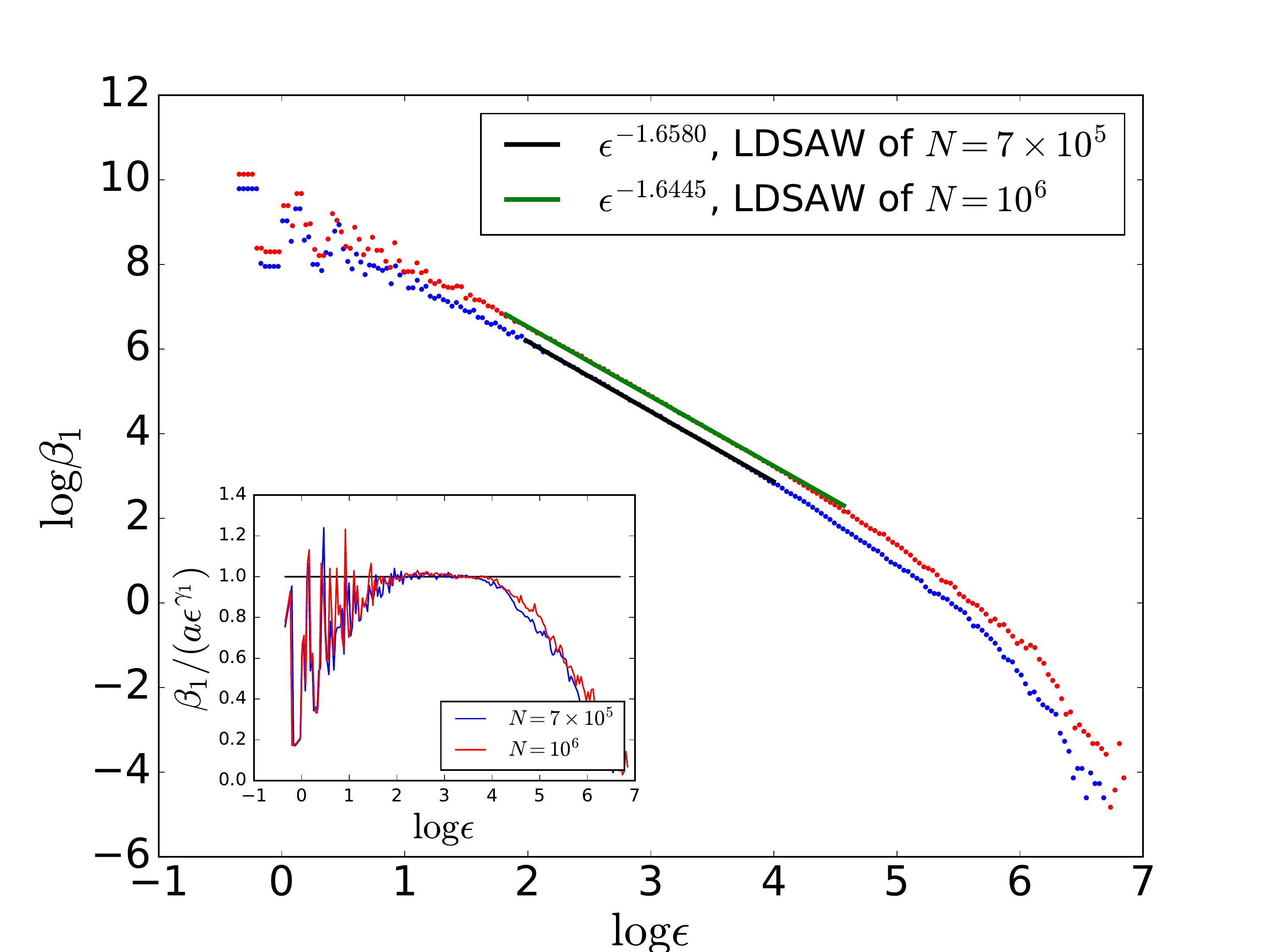}
\caption{The dependence of first Betti number on resolution $\epsilon$ for the LDSAW. The first Betti number is averaged over 200 independent conformations of LDSAW. Shown in 
the inset is the ratio between the Betti number and the power law relation.}
\label{fig:1stBettiLDSAW}
\end{figure}

The growth of the first Betti number of the LDSAW is shown in Fig.~\ref{fig:1stBettiLDSAW}. The estimated values of $\gamma_1$ for deleting the loops of SAW $N=7\times 10^5$ and $N=10^6$ 
are $1.6580$ and $1.6445$.

\subsection{The Geometric and Topological Properties of the Contacts of a Self-Avoiding Walk}
\label{sec:contact}
The contacts of inter- or intra-biomolecules often indicate the realization of some biological functions. These contacts 
are driven by many kinds of interactions such as electrostatic forces and hydrogen bonds. For a totally flexible linear 
chain the contacts are influenced only by the excluded volume interaction. The average number of contacts of self-avoiding walks 
is found to have an asymptotic behavior with respect to the length of the walks ($\langle N_c \rangle_{\text{SAW}} \sim 
a_{\infty} N$)~\cite{Douglas1995}. In 
this section, we investigate the distribution of the contacts from geometric and topological aspects.

\begin{figure}[hbt]\centering
\includegraphics[width=0.7\textwidth]{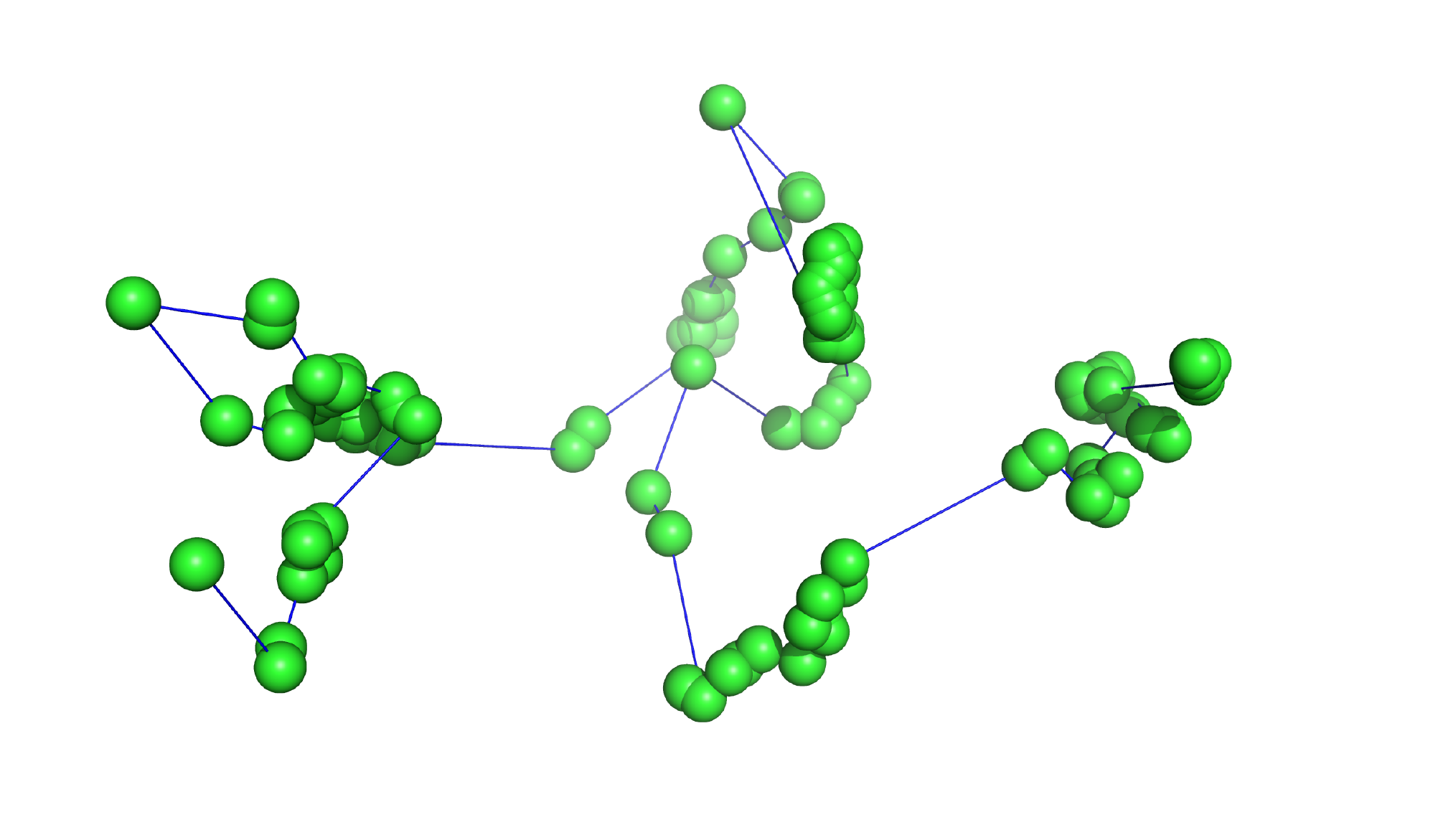}
\caption{The contacts of a self-avoiding walk. Each contact is represented by a green spheroid. The blue lines do not 
mean that the contacts are connected and just show the separation of different contacted regions. }
\label{fig:contact}
\end{figure}

In the cubic lattice model of the SAW, when two non-consecutive mono\-mers are nearest neighbors, 
they are 
counted as a contact, the position of which is the average of these two monomers. The set of those contact 
points is clearly near the backbone of the original SAW, that means it is a subset of the SAW. However, the fractality 
is not necessarily the same. Due to the excluded volume effect, two densely contacted regions should be separated 
spatially (Fig.~\ref{fig:contact}).

\begin{figure}[H]\centering
\includegraphics[width=0.7\textwidth]{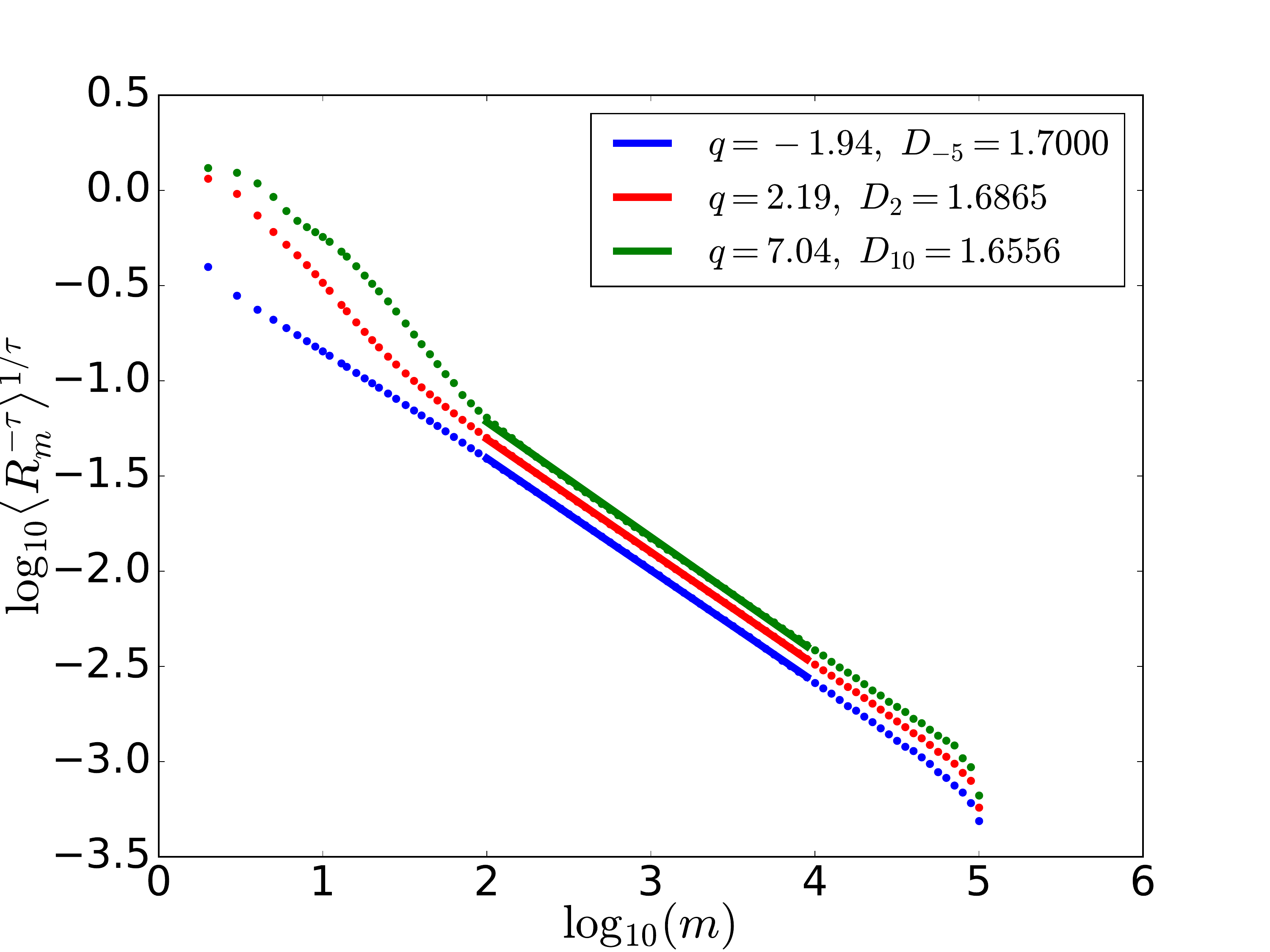}
\caption{Log-log plot of three moments of $R_m$ versus $m$ for the contacts of $N=10^6$ self-avoiding walk. The 
moments are $\tau=-5,\ 2,\ 10$.}
\label{fig:FDsawct}
\end{figure}

\begin{figure}[hbt]\centering
\includegraphics[width=0.7\textwidth]{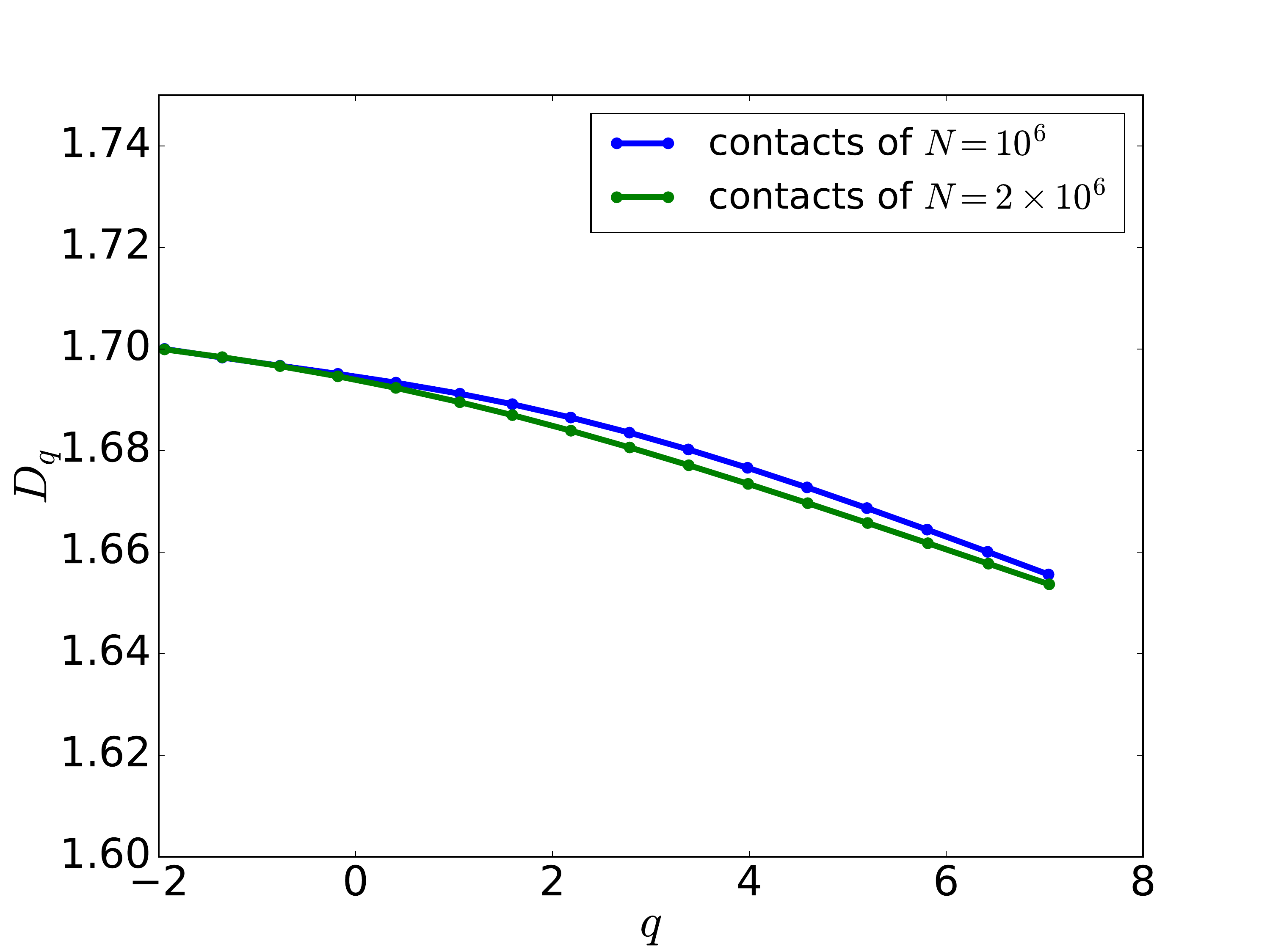}
\caption{The $D_q-q$ relation of the contact points within the SAW $N=10^6$ and $N=2\times 10^6$.}
\label{fig:multiFDsawct}
\end{figure}

Different from the self-avoiding walk, the point-cloud of contact seems to be disordered. 
To explore this, the Barycentric Fixed-Mass method is applied to study its fractality. The 
contacts within self-avoiding walks of $N=10^6$ and $N=2\times 10^6$ are recorded, each with 250 and 200 conformations. By 
calculating the average radius of $m$ nearest points (Eq.~(\ref{dq-mass})), we indeed find a power law dependence (Fig.~\ref{fig:FDsawct}) on $m$, 
which suggests that the set of contacts is fractal. However, $D_q$ exhibits a decreasing with the moment order 
$q$ (Fig.~\ref{fig:multiFDsawct}), which implies that the set of contacts may be multi-fractal.

\begin{figure}[hbt]\centering
\includegraphics[width=0.7\textwidth]{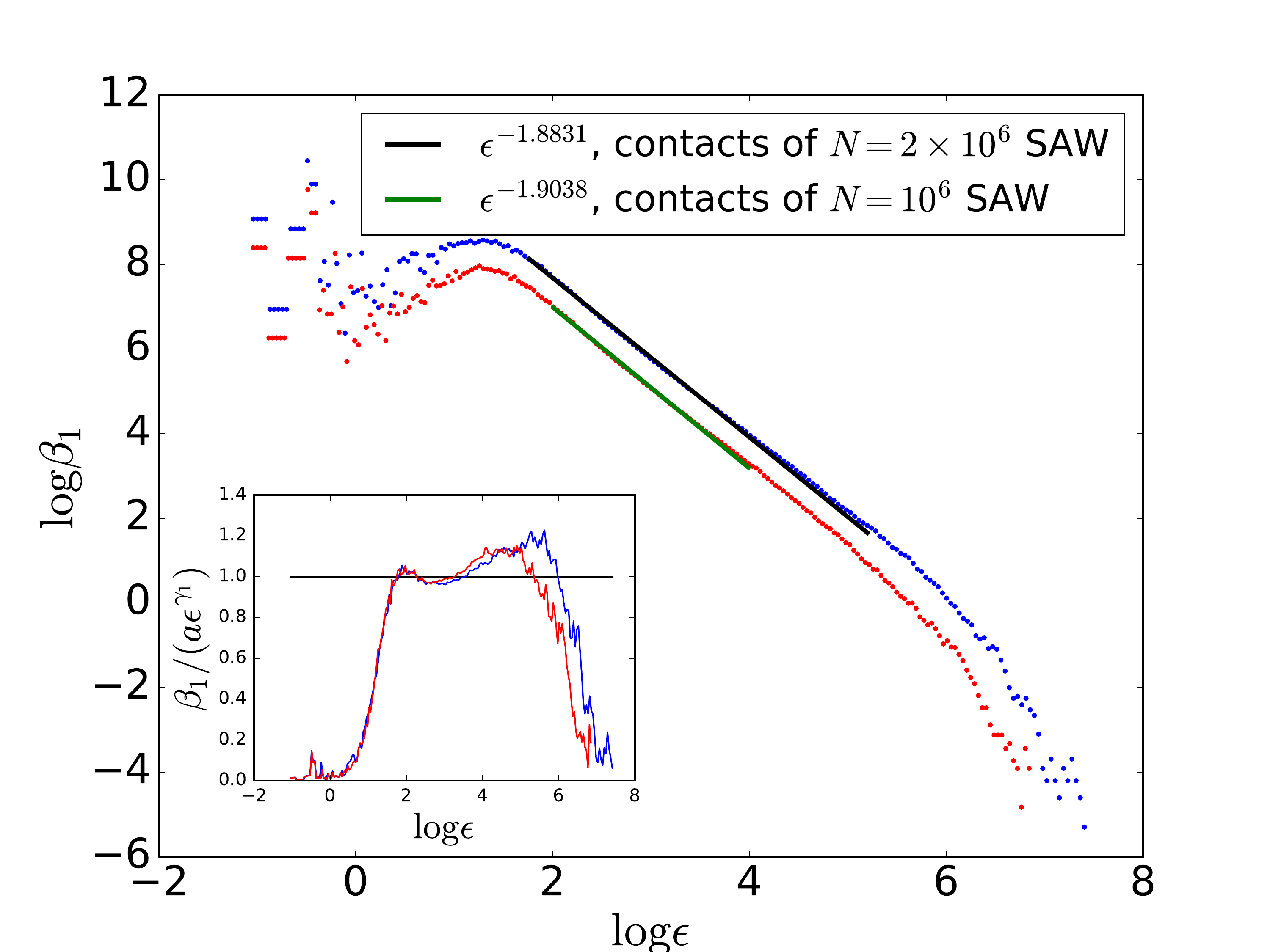}
\caption{The dependence of first Betti number on resolution $\epsilon$ for contacts within SAW of $N=2\times 10^6$ and 
$N=10^6$. The first Betti number is averaged over 200 and 250 independent conformations of the SAW. Shown in the inset 
is the ratio between the Betti number and the power law relation.}
\label{fig:1stBettiSAWct}
\end{figure}

The growth of the first Betti number of the contacts is shown in Fig.~\ref{fig:1stBettiSAWct}. The estimated values of $\gamma_1$ for the contacts of SAW $N=2\times 10^6$ and $N=10^6$ 
are $1.8831$ and $1.9038$.

\subsection{Randomly Deleting Points from a SAW}
\label{sec:rdsaw}
\begin{figure}[H]\centering
\includegraphics[width=0.7\textwidth]{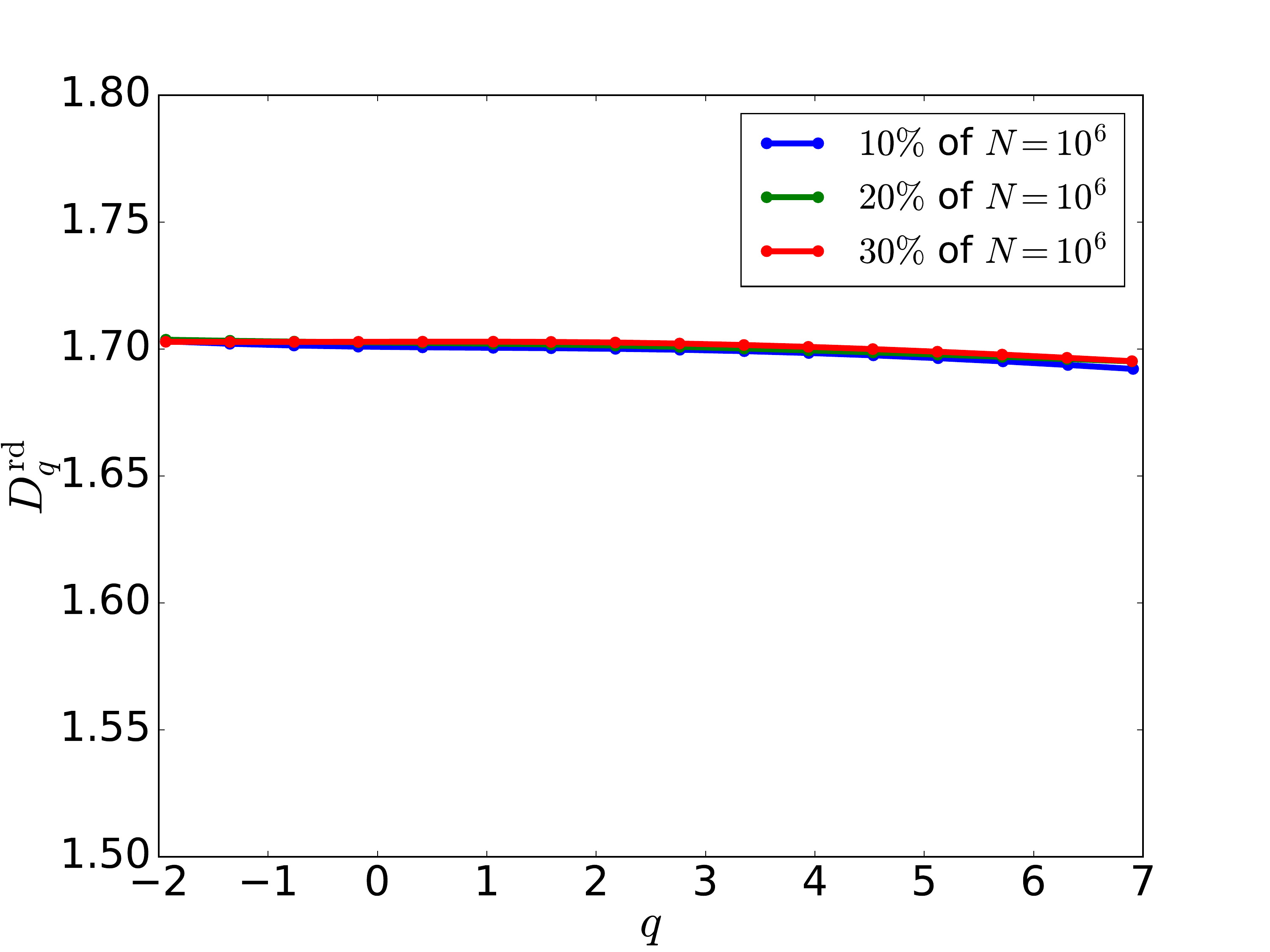}
\caption{$D(q)-q$ plot of three random subsets of the $N=10^6$ SAW using the Barycentric Fixed-Mass 
method. The three subsets contain 10\%, 20\%, 30\% of the points in the SAW respectively. The results are averaged over 300 
conformations.}
\label{fig:multiDSAWrd}
\end{figure}

Both the LDSAW in section~\ref{sec:LDSAW} and the contacts of the SAW~\ref{sec:contact} are subsets of the 
SAW, but they carry different geometrical and topological information. In this section, for comparison, we study another kind of
subset by deleting points in a SAW with equal probability. It is expected that this kind of subset has the same fractality as the SAW. In Fig.~\ref{fig:multiDSAWrd} is shown the multi-fractal analysis of three random 
subsets of the $N=10^6$ SAW using the Barycentric Fixed-Mass method. The three subsets contain 10\%, 20\%, 30\% of the points in the SAW respectively. The results are averaged over 300 conformations. It shows that this random subset has same 
fractal dimension as the SAW: it is mono-fractal with dimension $D^{\text{rd}}\approx 1.7$. 

\begin{figure}[H]\centering
\includegraphics[width=0.7\textwidth]{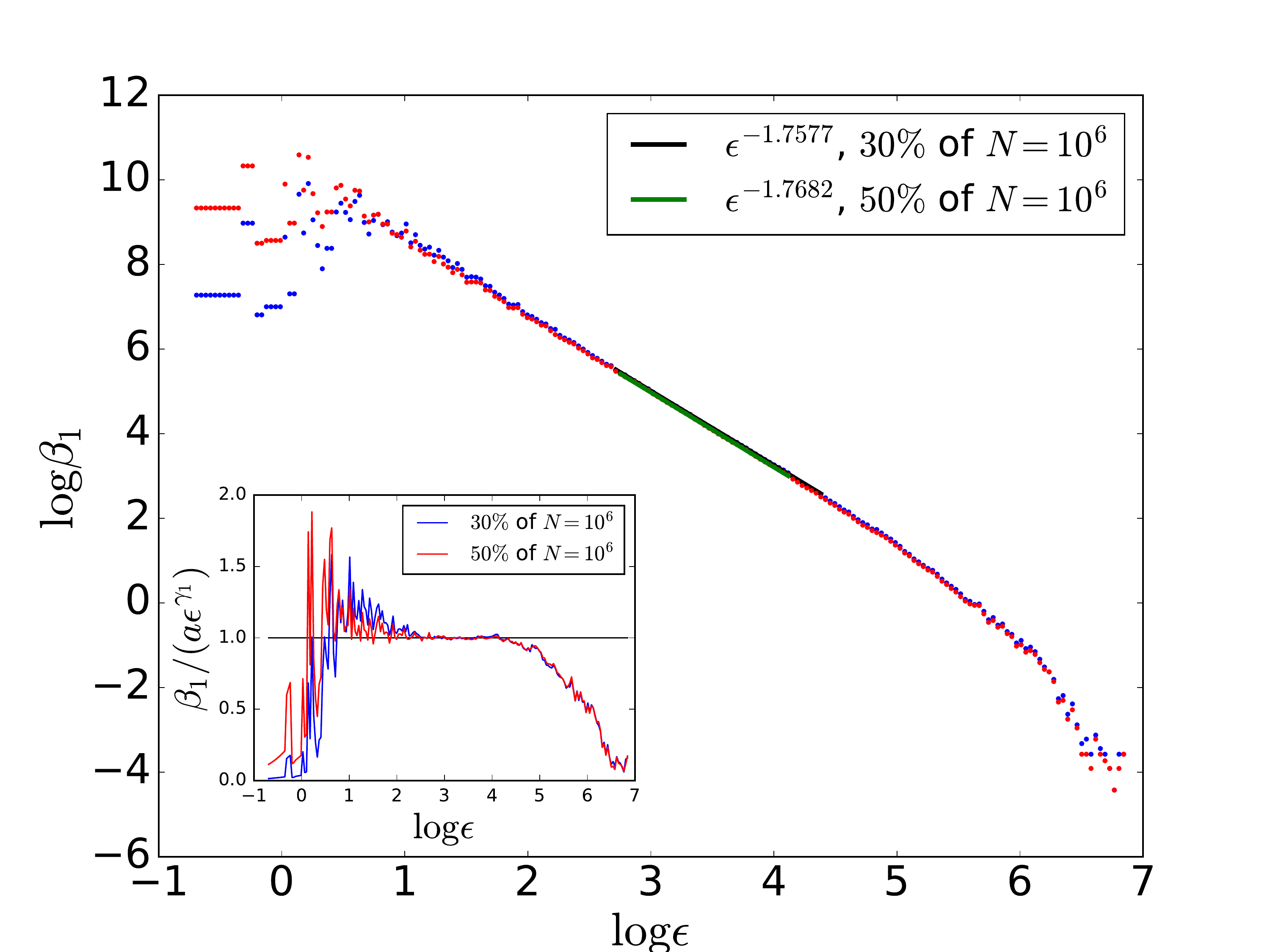}
\caption{The dependence of first Betti number on resolution $\epsilon$ for random subsets of a SAW. The 
first 
Betti number is averaged over $300$ independent conformations. Shown in the inset is the ratio between the Betti number 
and the power law relation.}
\label{fig:1stBettiSAWrd}
\end{figure}

The growth of the first  Betti number of the random subsets of the SAW is shown in Fig.~\ref{fig:1stBettiSAWrd}. The estimated values of $\gamma_1$ for 300 000 and 500 000 random points are $1.7577$ and $1.7682$.

\section{Conclusion}
\label{conclu}
We defined a new kind of walk: the loop-deleted self-avoiding walk in this paper. Its critical exponent was estimated in two ways as explained in section~\ref{sec:ldsawdef}, which arrives at the same result: $\lambda\approx 0.600$. We studied the difference of the point-cloud of the LDSAW and that of the SAW by calculating the fractal dimension and the growth rates of the Betti number. The fractal dimension is actually the reciprocal of the critical exponent, while the growth rate of the first Betti number is about 1.75 for the SAW, 1.65 for the LDSAW. The spatial distribution of the contacts inside a SAW is also analyzed, with the contact-cloud also being a subset of the SAW point-cloud. The result shows that the contact-cloud is multi-fractal, while the growth rate of the first Betti number is about 1.90. Finally, for comparison, we study the random subset of a SAW. This random subset has the same fractal dimension and a close growth rate of the first Betti number with the SAW.

Clearly, the properties of a subset of the SAW point-cloud hinge on how it is extracted, which leads to the LDSAW, contact-cloud, and random deleting subset discussed here. We calculated their fractal dimension and the growth rates of the Betti number. Till now there is no theoretical studies of the growth behavior of the Betti numbers for the SAW, but it is assumed that the growth rates are related to the fractal dimension according to Robins~\cite{Robins:2000} which is still not fully understood.

\section*{Acknowledgments}
DWH would like
to express his sincere thanks to Dietrich Stauffer. Without his guidance the trajectory of life would have been a completely different one!
JJ would like to
acknowledge funding from the China Scholarship Council (CSC NO.201506210082). 
Part of this work was funded by the Deutsche Forschungsgemeinschaft (DFG, German Research Foundation) under Germany's Excellence Strategy EXC-2181/1 - 390900948.

\end{document}